\documentclass[12pt]{article}
\usepackage{amssymb}
\usepackage{epsfig}
\voffset= - 0.7in
\hoffset= - 0.6in         

\catcode`\@=11
\def\marginnote#1{}

\newcount\hour
\newcount\minute
\newtoks\amorpm
\hour=\time\divide\hour by60
\minute=\time{\multiply\hour by60 \global\advance\minute by-\hour}
\edef\standardtime{{\ifnum\hour<12 \global\amorpm={am}%
        \else\global\amorpm={pm}\advance\hour by-12 \fi
        \ifnum\hour=0 \hour=12 \fi
        \number\hour:\ifnum\minute<10 0\fi\number\minute\the\amorpm}}
\edef\militarytime{\number\hour:\ifnum\minute<10 0\fi\number\minute}

\def\draftlabel#1{{\@bsphack\if@filesw {\let\thepage\relax
   \xdef\@gtempa{\write\@auxout{\string
      \newlabel{#1}{{\@currentlabel}{\thepage}}}}}\@gtempa
   \if@nobreak \ifvmode\nobreak\fi\fi\fi\@esphack}
        \gdef\@eqnlabel{#1}}
\def\@eqnlabel{}
\def\@vacuum{}
\def\draftmarginnote#1{\marginpar{\raggedright\scriptsize\tt#1}}

\def\draft{\oddsidemargin -.5truein
        \def\@oddfoot{\sl preliminary draft \hfil
        \rm\thepage\hfil\sl\today\quad\militarytime}
        \let\@evenfoot\@oddfoot \overfullrule 3pt
        \let\label=\draftlabel
        \let\marginnote=\draftmarginnote
   \def\@eqnnum{(\theequation)\rlap{\kern\marginparsep\tt\@eqnlabel}%
\global\let\@eqnlabel\@vacuum}  }

\def\appname{Appendix}
\newcounter{app}
\def\theapp{\Alph{app}}
\def\app{\par
   \addvspace{4ex}
   \@afterindentfalse
  \secdef\@app\@dapp}
\def\@app[#1]#2{\ifnum \c@secnumdepth >\m@ne
        \refstepcounter{app}
        \addcontentsline{toc}{app}{\theapp
        \hspace{1em}#1}\else
      \addcontentsline{toc}{app}{ #1}\fi
   {\parindent \z@ \raggedright
    \Large \bf \appname~\theapp .
   \Large  \bf 
    #2}\nobreak
   \vskip 4ex   \noindent
\setcounter{equation}{0}
\def\theequation{\Alph{app}.\arabic{equation}}}
\def\@dapp#1{%
{\parindent \z@ \raggedright  \bf #1}\par\nobreak}
\def\l@app#1#2{\addpenalty{\@secpenalty}%
   \addvspace{1em plus\p@}%
   \begingroup
   \@tempdima 3em
     \parindent \z@ \rightskip \@pnumwidth
     \parfillskip -\@pnumwidth
     { \bf
     \leavevmode
     #1\hfil \hbox to\@pnumwidth{\hss #2}}\par
     \nobreak
   \endgroup}

\makeatletter
\newdimen\normalarrayskip            
\newdimen\minarrayskip               
\normalarrayskip\baselineskip
\minarrayskip\jot
\newif\ifold             \oldtrue            \def\new{\oldfalse}
\def\arraymode{\ifold\relax\else\displaystyle\fi}
\def\eqnumphantom{\phantom{(\theequation)}} 
\def\@arrayskip{\ifold\baselineskip\z@\lineskip\z@
     \else
     \baselineskip\minarrayskip\lineskip1\baselineskip\fi}

\def\@arrayclassz{\ifcase \@lastchclass \@acolampacol \or
\@ampacol \or \or \or \@addamp \or
   \@acolampacol \or \@firstampfalse \@acol \fi
\edef\@preamble{\@preamble
  \ifcase \@chnum
     \hfil$\relax\arraymode\@sharp$\hfil
     \or $\relax\arraymode\@sharp$\hfil
     \or \hfil$\relax\arraymode\@sharp$\fi}}

\def\@array[#1]#2{\setbox\@arstrutbox=\hbox{\vrule
     height\arraystretch \ht\strutbox
     depth\arraystretch \dp\strutbox
width\z@}\@mkpream{#2}\edef\@preamble{\halign \noexpand\@halignto
\bgroup \tabskip\z@ \@arstrut \@preamble \tabskip\z@ \cr}%
\let\@startpbox\@@startpbox \let\@endpbox\@@endpbox
  \if #1t\vtop \else \if#1b\vbox \else \vcenter \fi\fi
  \bgroup \let\par\relax
  \let\@sharp##\let\protect\relax
  \@arrayskip\@preamble}
\def\eqnarray{\stepcounter{equation}%
              \let\@currentlabel=\theequation
              \global\@eqnswtrue
              \global\@eqcnt\z@
              \tabskip\@centering              
              \let\\=\@eqncr
              $$%
            \halign to \displaywidth  \bgroup
             \eqnumphantom \@eqnsel
      \hskip\@centering                               
    $\displaystyle  \tabskip\z@ {##}$%
    &\global\@eqcnt\@ne \hskip 2\arraycolsep
         $ \displaystyle  \arraymode{##}$\hfil
    &\global\@eqcnt\tw@ \hskip 2\arraycolsep
         $\displaystyle\tabskip\z@{##}$\hfil
         \tabskip\@centering
    &{##}\tabskip\z@\cr}
\makeatother

\newfont{\hr}{msbm10}
\newfont{\ams}{msam10}

\font\numbers=cmss12
\font\upright=cmu10 scaled\magstep1
\def\stroke{\vrule height8pt width0.4pt depth-0.1pt}
\def\topfleck{\vrule height8pt width0.5pt depth-5.9pt}
\def\botfleck{\vrule height2pt width0.5pt depth0.1pt}
\def\Zmath{\vcenter{\hbox{\numbers\rlap{\rlap{Z}\kern 0.8pt\topfleck}\kern
2.2pt
                   \rlap Z\kern 6pt\botfleck\kern 1pt}}}
\def\Qmath{\vcenter{\hbox{\upright\rlap{\rlap{Q}\kern
                   3.8pt\stroke}\phantom{Q}}}}
\def\Nmath{\vcenter{\hbox{\upright\rlap{I}\kern 1.7pt N}}}
\def\Cmath{\vcenter{\hbox{\upright\rlap{\rlap{C}\kern
                   3.8pt\stroke}\phantom{C}}}}
\def\Rmath{\vcenter{\hbox{\upright\rlap{I}\kern 1.7pt R}}}
\def\Z{\ifmmode\Zmath\else$\Zmath$\fi}
\def\Q{\ifmmode\Qmath\else$\Qmath$\fi}
\def\N{\ifmmode\Nmath\else$\Nmath$\fi}
\def\C{\ifmmode\Cmath\else$\Cmath$\fi}
\def\R{\ifmmode\Rmath\else$\Rmath$\fi}

\def\d{\partial}

\def\bea{\begin{eqnarray}}
\def\eea{\end{eqnarray}}

\def\beq{\begin{equation}}
\def\eeq{\end{equation}}
\def\ba{\beq\new\begin{array}{c}}
\def\ea{\end{array}\eeq}
\def\be{\ba}
\def\ee{\ea}

\def\e{\epsilon}

\def\stackreb#1#2{\mathrel{\mathop{#2}\limits_{#1}}}
\def\Tr{{\rm Tr}}

\def\half{{\textstyle{1\over2}}}
\def\ha{{1\over 2}}
\def\N2{${\cal N}=2$}
\def\4N{${\cal N}=4$}
\def\1N{${\cal N}=1$}
\def\1N*{${\cal N}=1^*$}

\def\beq{\begin{equation}}
\def\eeq{\end{equation}}
\def\ba{\beq\new\begin{array}{c}}
\def\ea{\end{array}\eeq}
\def\be{\ba}
\def\ee{\ea}
\newcommand{\rf}[1]{(\ref{#1})}

\begin{document}


\begin{flushright}
FIAN/TD-09/09\\
ITEP/TH-26/09
\end{flushright}
\vspace{1.0 cm}

\setcounter{footnote}{3}
\renewcommand{\thefootnote}{\fnsymbol{footnote}}
\begin{center}
{\Large\bf
First Order String Theory and\\
\vspace{0.3 cm}
the Kodaira-Spencer Equations. II
}\\
\vspace{1.5 cm}
{\large O.~Gamayun\footnote{
Bogolyubov Institute of Theoretical Physics, Kiev, Ukraine} and
A.~Marshakov\footnote{
Lebedev Physics Institute and
Institute for Theoretical and Experimental Physics, Moscow, Russia}}\\
\vspace{0.6 cm}
\end{center}
\vspace{0.3 cm}
\begin{quotation}
\noindent
The first-order bosonic string theory, perturbed by primary
operator, corresponding to the deformation of target-space complex structure
is considered. We compute the correlation functions in this theory and study
their divergencies. It is found, that consistency of these correlation functions
with the world-sheet
conformal invariance requires the Kodaira-Spencer equations to be satisfied
by target-space Beltrami differentials. This statement is checked explicitly
for the three-point and four-point correlators, containing one probe operator.
We discuss the origin of these divergences and their relation with
beta-functions or effective action and polyvertex structures in BRST approach.
\end{quotation}
\renewcommand{\thefootnote}{\arabic{footnote}}
\setcounter{section}{0} \setcounter{footnote}{0}
\setcounter{equation}0

\section{Introduction}

This paper continues our study of the first order bosonic string theory and completes
the results of \cite{GLM1}. We propose the formulation of string theory
in non-trivial background, to be hopefully an alternative to traditional approach
(see e.g. \cite{Pol,GSW,FT,CFoth}), based
almost totally on studying the two-dimensional sigma-models. The proposed in
\cite{LMZ,GLM1} first-order string theory is based on the perturbation of the ``bare action''
for (coupled to ghosts) bosonic first-order free conformal theory
\be
\label{act}
S_0=\frac{1}{2\pi\alpha'}\int_\Sigma d^2 z (p_i\bar{\partial} X^{i}+
p_{\bar{i}}{\partial} X^{\bar{i}})
\ee
which is independent of the target-space metric and requires only some local choice
of the target-space complex structure. The world-sheet
fields $\{ X^\mu \} = \{\left(X^i,X^{\bar i}\right)\}$,
$\{ p_i \}$ and $\{ p_{\bar i} \}$ (with $\mu=1,\ldots,D$; $i,{\bar i}=1,\ldots,D/2$) are
sections of $H^0(\Sigma)$, $H^{(1,0)}(\Sigma)$ and $H^{(0,1)}(\Sigma)$ correspondingly,
being holomorphic
(or anti-holomorphic) on the equations of motion. The only nontrivial operator product
expansions (OPE) for the theory \rf{act} are
\be
\label{px}
p_i(z) X^j(z') = \frac{\alpha'\delta_i^j}{z-z'} + {\rm regular\ terms}
\ee
together with their complex conjugated, and computation of all nontrivial
correlation functions on sphere in the theory \rf{act} is therefore reduced to collection of
correlators \rf{px} by application of the Wick theorem.

The free field theory action \rf{act} can be naturally perturbed by the operators
\be
\label{vg}
V_g = \frac{1}{2\pi\alpha'}\int_\Sigma O_g =
\frac{1}{2\pi\alpha'}\int_\Sigma d^2z g^{i\bar{j}}p_i p_{\bar{j}}
\ee
with the $X$-dependent ``coefficient functions'' or target-space fields $g^{i\bar{j}} =
g^{i\bar{j}}(X)$, as well as
\be
\label{vmu}
V_\mu = \frac{1}{2\pi\alpha'}\int_\Sigma O_\mu =
\frac{1}{2\pi\alpha'}\int_\Sigma d^2z \mu_{\bar{i}}^j\bar{\partial} X^{\bar{i}} p_j
\\
V_{\bar\mu} = \frac{1}{2\pi\alpha'}\int_\Sigma O_{\bar\mu} =
\frac{1}{2\pi\alpha'}\int_\Sigma d^2z \bar{\mu}^{\bar{j}}_i\partial X^i p_{\bar{j}}
\ee
where $\mu_{\bar{i}}^j = \mu_{\bar{i}}^j(X)$ (together with its complex conjugated
${\bar\mu}^{\bar{i}}_j = {\bar\mu}^{\bar{i}}_j(X)$), and
\be
\label{vb}
V_b = \frac{1}{2\pi\alpha'}\int_\Sigma O_b =
\frac{1}{2\pi\alpha'}\int_\Sigma d^2z b_{i\bar{j}}\partial X^i\bar{\partial} X^{\bar{j}}
\ee
where again $b_{i\bar{j}} = b_{i\bar{j}}(X)$. We shall use
the ``real'' operator
\be
\label{Fimu}
\Phi(z,{\bar z}) = O_\mu(z,{\bar z}) + O_{\bar\mu}(z,{\bar z}) =
\mu_{\bar{i}}^j\bar{\partial} X^{\bar{i}} p_j
+ \bar{\mu}^{\bar{j}}_i\partial X^i p_{\bar{j}}
\ee
In order for the operators \rf{vg},\rf{vmu} and \rf{Fimu} to be well-defined as
conformal primary operators, one has to impose the transversality conditions for
the background fields
\be
\label{transv}
\d_i g^{i{\bar j}} = 0,\ \ \ \d_{\bar j} g^{i{\bar j}} =0
\\
\d_i \mu_{\bar{j}}^i = 0,\ \ \ \d_{\bar j} \bar{\mu}^{\bar{j}}_i  =0
\ee
which allow to get rid of the singularities, possibly arising from ``internal'' contractions in
\rf{vg},\rf{vmu} and \rf{Fimu} or, in different words, the higher-order poles
in the operator-product expansions with the components of the
stress-energy tensor $T \sim p_i\d X^i$ and ${\bar T} \sim p_{\bar i}{\bar\d} X^{\bar i}$ in the
bare theory \rf{act}. In the BRST approach, to be also briefly discussed below, conditions \rf{transv} follow directly from requiring the operators \rf{vg} and \rf{vmu} to be BRST-closed.

The operators \rf{vg}-\rf{vb} (or \rf{Fimu}) are the only possible
marginal $(\Delta,{\bar\Delta})=(1,1)$
primary operators in the first-order theory \rf{act}. In addition, one can also introduce the
holomorphic $(1,0)$-currents
\be
\label{cur}
j_v = p_iv^i(X),\ \ \ \d_iv^i=0
\\
j_\omega = \omega_i(X)\d X^i
\ee
(and their anti-holomorphic $(0,1)$-conjugates),
which generate the holomorphic change of co-ordinates and gauge transformations (their
anomalous operator algebra has been studied in \cite{MSV,W,LMZ,N}). The non-holomorphic
operators, similar to \rf{cur}, can also arise when studying generic non-holomorphic
symmetries of the perturbed action \cite{GLM1}. We shall also see, how such structures arise
in the operator algebra of the background operators \rf{vg}-\rf{vb}
and in the extra divergences, related to the anomalies of the currents \rf{cur}.

Below we
are going to find the conditions, when the operators \rf{vg}-\rf{vb} become {\em exactly}
marginal or can be raised up to the exponent and added to the free action \rf{act}.
In other words, this is equivalent to vanishing of their beta-functions
in the perturbed theory \cite{Pol,Zam,GLM1}.
The quadratic (in background fields) contributions to these beta-functions are
given by the structure constants of the OPE's of the primary operators \rf{vg}-\rf{vb}
(see Appendix~\ref{ap:ope}),
whose vanishing leads, for example, to the nonlinear equation \cite{LMZ}
\be
\label{geq}
g^{i\bar{j}}\d_i\d_{\bar j}g^{k\bar{l}}-\d_ig^{k\bar{j}}\d_{\bar j}g^{i\bar{l}}=0
\ee
for the functions $g^{i\bar{j}}(X)$. Since the background field equations
are generally highly nonlinear, it is clear, that
exact form of the beta-functions should be affected by certain
{\em polyvertex} contributions, when more than
two vertex operators collide on world sheets.
In order to get such contributions explicitly, one needs to study the
effective action beyond quadratic level, or the multi-point correlation functions
in the first-order theory \rf{act}.

In the present paper,
we would like to concentrate mostly on the background equations of motion
for the target-space
``Beltrami'' fields\footnote{
These fields in the context of Lagrangian field theory were discussed already in
\cite{Lazar}, the Beltrami parametrization of the world-sheet geometry in string theory
was discussed e.g. in \cite{Baul}.} $\mu = dX^{\bar{j}}\mu_{\bar{j}}^i{\d \over\d X^i}$
and $\bar\mu = dX^i \bar{\mu}^{\bar{j}}_i{\d \over\d X^{\bar{j}}}$, keeping the other fields
to be shut down for a while, or playing maximally a role of a ``spectator'' or
``probe'' operators.
In such case the vertex operators \rf{vmu} and \rf{Fimu} can be obviously considered as
deforming the complex structure of the original bare theory \rf{act}, and from generic
target-space symmetry reasons one would expect that the corresponding fields should
satisfy the Kodaira-Spencer equations \cite{KoSpe}
\be
\label{KoSpe}
N_{{\bar k}{\bar j}}^i \equiv \d_{[{\bar k}}\mu^i_{{\bar j}]} -
\mu_{[\bar k}^l\d_l\mu_{\bar j]}^i =0
\\
{\bar N}_{ik}^{\bar j} \equiv \d_{[i}{\bar\mu}^{\bar j}_{k]} -
{\bar\mu}_{[i}^{\bar l}\d_{\bar l}{\bar\mu}_{k]}^{\bar j} = 0
\ee
which have an obvious sense of vanishing of the Nijenhuis tensor or
curvatures for the gauge fields
$\mu = dX^{\bar{j}}\mu_{\bar{j}}^i{\d \over\d X^i}$ and
$\bar\mu = dX^i \bar{\mu}^{\bar{j}}_i{\d \over\d X^{\bar{j}}}$ with the values in Lie algebra
of the vector fields in tangent bundle to the target manifold (see \cite{BCOV} for brief
description of Kodaira-Spencer theory and their important applications for topological
strings). Below we are going to derive these equations directly from
computation of the correlation functions in the first-order conformal field theory.

Let us immediately point out the most intriguing features and attractive outcomes of solving
this problem:
\begin{itemize}
  \item The Kodaira-Spencer equations \rf{KoSpe} are {\em nonlinear} and, in contrast to
  \rf{geq} contain terms of different powers in the background Beltrami fields.
  From the point of view
  of world-sheet theory it means that they result from considering not just the simplest singularities of the correlators, or the OPE's, but rather from the higher singularities of the multipoint correlation functions with different numbers of the
  entries. In particular, it allows to test some general hypothesis about
  relevance of the polyvertex
  structures in string theory, related to nontrivial boundary components of the multipoint moduli
  spaces ${\cal M}_{g,n}$ of the world-sheet curves.
\item Below we are going to compute the correlation functions in the first-order
  conformal field theory, and study their (logarithmic) divergencies.
  We show, that vanishing of the corresponding target-space coefficient functions
  correspond rather to a certain bilinear combination of the Kodaira-Spencer
  equations \rf{KoSpe}. The exact form of this combination was determined in \cite{GLM1}
  from standard
  computation of the effective action, or more strictly - of the beta-function for the
  operator \rf{vb}. We demonstrate now, how the quadratic and cubic pieces of this
  expression can be extracted from direct computation of the 3-point and 4-point
  correlation functions. We compute these contributions from the co-ordinate
  representations for these correlators, making a first
  step towards the systematic study of the co-ordinate beta-functions as integrals over the
  world-sheet moduli spaces in string theory, and we discuss also extra
  singularities of these correlators. Finally, we turn to the interpretation of
  the computed beta-functions within the BRST approach and show how our results can be
  partially reproduced in terms of generalized Maurer-Cartan equation.

\end{itemize}

\section{The perturbed correlation functions}

To study the co-ordinate approach to the beta-functions consider, for example, the perturbed
one-point correlation function of the ``probe operator'' \rf{vg}
\be
\label{gfi}
\langle O_g(x)\rangle_t = \langle O_g(x)\exp (t\int_\Sigma\Phi)\rangle =
\sum_{n\geq 0}{t^n\over n!}\int_\Sigma d^2z_1\ldots\int_\Sigma d^2z_n
\langle  O_g(x)\Phi(z_1)\ldots\Phi(z_n)\rangle
\ee
where averaging in the r.h.s. is understood in the sense of path integral with the free
action \rf{act}. The calculation of the r.h.s. of \rf{gfi} includes the integration of
the multipoint correlators
\be
\label{corint}
\langle O_g(x)\Phi(z_1)\ldots\Phi(z_n)\rangle
\ee
over the regularized domain $\Sigma^{\otimes n}$, e.g.
\be
\label{dom}
|z_i-x|>\epsilon,\ \ \ \ |z_i-z_j|>\epsilon
\\
\forall i,j=1,\ldots,n
\ee
To study the boundaries of the moduli spaces in detail, one should in principle consider
{\em different} $\epsilon$'s, or even to take $\epsilon=\epsilon(z,{\bar z})$
in these inequalities.
It is also sometimes useful to introduce the IR cutoff $R$ and discuss the correlators
\rf{corint} in the regime, when $|z_i|\ll R$, $\forall i$, while $|x| \gg R$, what corresponds
to extracting the UV beta-function divergences from \rf{gfi}.

We shall be interested in what follows only in the terms arising at
{\em logarithmic} in UV cutoff $\epsilon$ singularities,
when integrating the correlation functions \rf{corint}. We are going to demonstrate, that vanishing
of such terms gives, in particular, the expected from alternative approach \cite{GLM1} contributions to the squared Kodaira-Spencer
equations \rf{betab}, imposed onto the set of target-space
Beltrami differentials $\mu$ and ${\bar\mu}$. It is essential, that the Kodaira-Spencer equations
are nonlinear, and in order to get them one should carefully take into account the contribution
of different multipoint functions from the set \rf{corint}.

\subsection{3-point function
\label{ss:3pt}}

In order to get a hint of what should we expect from such computation, consider the first
nontrivial order, namely the correlator \rf{corint} for $n=2$
\be
\label{gfifi}
\langle O_g(x)\Phi(y)\Phi(z)\rangle =
\langle O_g(x)O_\mu(y) O_{\bar\mu}(z)\rangle +
\langle O_g(x)O_{\bar\mu}(y) O_\mu(z)\rangle
\ee
The direct computation of the free field correlator in the r.h.s. of \rf{gfifi}
gives rise to the result
\be
\label{gfifir}
C_2=\ha\langle O_g(x)\Phi(y)\int_\Sigma d^2z \Phi(z)\rangle =
{g^{i{\bar j}}B^{(2)}_{i{\bar j}}\over|x-y|^2}
\int_\Sigma {d^2z\over\pi}{1\over |x-z|^2|y-z|^2}
\ee
The integral in \rf{gfifir}, which is basically the ``volume'' of the group $SL(2,\mathbb{C})$
of global transformations on sphere, when computed over the domain
\rf{dom} $D_\epsilon(x,y) =\{|z-x|>\epsilon, |z-y|>\epsilon\}\subset\mathbb{C}$ in the complex plane
contains the logarithmic divergence (for the details of the calculation of the integrals
see Appendix~\ref{ap:int})
\be
\label{intlog}
\int_{D_\epsilon(x,y)} {d^2z\over\pi}{1\over |x-z|^2|y-z|^2}\simeq
{2\over |x-y|^2}\log{|x-y|^2\over\epsilon^2}
\ee
The rest integration over $y$, omitted in \rf{gfifir} is not essential: if introducing
the IR cutoff the integrated singularity \rf{intlog} can be just replaced by
\be\label{C2}
C_2 \approx \frac{g^{i\overline{j}}B^{(2)}_{i\overline{j}}}{|x|^4}\log\frac{R^2}{\e^2}
\int\limits_{|y|<R} \frac{d^2y}{\pi}
\ee
The coefficient at the logarithmic singularity \rf{gfifir}, \rf{intlog}, \rf{C2}
is proportional to the function
\be
\label{ks0}
g^{i{\bar j}}B^{(2)}_{i{\bar j}} =
g^{k{\bar k}}\d_{[i}{\bar\mu}^{\bar j}_{k]}\d_{[{\bar k}}\mu^i_{{\bar j}]}
\ee
which has an obvious sense of the squared linearized Kodaira-Spencer equations \rf{KoSpe}.
The structure \rf{ks0} arises by straightforward direct computation of the free-field
correlation functions coming from \rf{gfifir}
\be
\label{ffcor}
\langle O_g(x)O_{\bar\mu}(y)O_\mu(z)\rangle =
{\langle g^{i\bar j}(x)\ p_{\bar k}{\bar\mu}_i^{\bar k}(y)\ p_k\mu_{\bar j}^k(z)\rangle
\over (x-y)^2({\bar x}-{\bar z})^2} +
{\langle g^{i\bar j}p_{\bar j}(x)\ {\bar\mu}_i^{\bar k}(y)\ p_k\mu_{\bar k}^k(z)\rangle
\over (x-y)^2({\bar y}-{\bar z})^2} +
\\
+ {\langle g^{i\bar j}p_ip_{\bar j}(x)\ {\bar\mu}_k^{\bar k}(y)\ \mu_{\bar k}^k(z)\rangle
\over (y-z)^2({\bar y}-{\bar z})^2} +
{\langle g^{i\bar j}p_i(x)\ p_{\bar k}{\bar\mu}_k^{\bar k}(y)\ \mu_{\bar j}^k(z)\rangle
\over (y-z)^2({\bar x}-{\bar z})^2}
\ee
due to vanishing of the one-point functions of quantum fields
$\langle\d X\rangle = \langle{\bar\d} X\rangle = 0$. Computing further the r.h.s. of \rf{ffcor}
one has to take into account the transversality \rf{transv} and the rule, allowing
integration by parts over the target-space zero
modes.

Strictly speaking, instead of
$g^{k{\bar k}}\d_{[i}{\bar\mu}^{\bar j}_{k]}\d_{[{\bar k}}\mu^i_{{\bar j}]}$ one should write
a target-space integral
\be
\label{tarint}
\int d^D X^{(0)}g^{k{\bar k}}(X^{(0)})\d_{[i}{\bar\mu}^{\bar j}_{k]}(X^{(0)})
\d_{[{\bar k}}\mu^i_{{\bar j}]}(X^{(0)})
\ee
over the zero modes $X^{(0)}$ of the $X$-fields,
and consider the compact target, or the target-space fields being
coefficients functions of the operators
\rf{vg} and \rf{vmu} vanishing at the space-time ``infinity'', what allows integration
by parts in \rf{tarint}. Together with the transversality constraints
\rf{transv} this bring us to \rf{gfifi} and its generalizations below.
The computation of the 3-point function \rf{gfifi} is almost equivalent to the
calculation of the operator product expansion of two operators $\Phi$ (see Appendix~\ref{ap:ope}
for details) with its further projection onto the operator $O_g$, being in this sense
equivalent to the computation of the quadratic contribution into the beta-function
of the operator $O_b$ \cite{GLM1}.
The integration by parts and transversality allows however to drop off the total derivatives
in the OPE \rf{OOPE},
and therefore the coefficient in front of the logarithmic singularity in \rf{gfifir} gives exactly
the desired term \rf{ks0}.

The singularity \rf{C2} can be compensated by the $B_{i\bar{j}}$-type counterterm
\be
\delta O_b^{(1)}
=\log\frac{\e^2}{\mu^2}B^{(2)}_{i\bar{j}}\d X^i\bar{\d} X^{\bar{j}}
\ee
so that the renormalized correlator becomes UV-finite in this order
\be
\langle O_g(x)\int
\frac{d^2z}{\pi}\delta O_b^{(1)}(z)\rangle + \frac{1}{2}\langle O_g(x)\int
d^2z_1 \Phi(z_1)\int
d^2z_2 \Phi(z_{2})\rangle \sim \frac{g^{i\bar{j}}B^{(2)}_{i\bar{j}}}{|x|^4}\log\frac{R^2}{\mu^2}
\int\limits_{|z|<R} \frac{d^2z}{\pi}
\ee
Since Kodaira-Spencer equations \rf{KoSpe} are nonlinear, the natural question now is whether
the multipoint correlation functions complete the expression $B^{(2)}_{i{\bar j}}$ in \rf{ks0}
to the squared Kodaira-Spencer equations in their exact form.
It is quite instructive to discuss now, how the next order $B^{(3)}_{i{\bar j}}$ arises from the
four-point contribution. This is already not very trivial computation, requiring special care,
when considering the integrals over the world-sheet moduli space, and we consider it in next section.

\subsection{4-point function: the free field correlator
\label{ss:4pt}}

In the next order for \rf{gfi}, \rf{corint} one gets the following contributions
\be
\label{gfi4}
\langle O_g(x)\Phi(y) \Phi(z)\Phi(w)\rangle =
\langle O_g(x)O_{\bar\mu}(y)\Phi(z)\Phi(w)\rangle + c.c. =
\langle O_g(x)O_{\bar\mu}(y) O_\mu(z) O_\mu(w)\rangle +
\\
+\langle O_g(x)O_{\bar\mu}(y) O_{\bar\mu}(z) O_\mu(w)\rangle+
\langle O_g(x)O_{\bar\mu}(y) O_\mu(z) O_{\bar\mu}(w) \rangle + c.c.
\ee
which now reduces to computation and integration \rf{gfi} of the free-field correlation functions
\be
\label{gmmm}
\langle O_g(x)O_{\bar\mu}(y)O_\mu(z)O_\mu(w)\rangle =
{\langle g^{i{\bar j}}p_i(x)\ \bar\mu_k^{\bar k}\d X^k(y)\
\mu_{\bar j}^l p_l(z)\ \mu_{\bar k}^r p_r(w)\rangle
\over ({\bar x}-{\bar z})^2({\bar y}-{\bar w})^2} +
\\
+ {\langle g^{i{\bar j}}p_i(x)\ \bar\mu_k^{\bar k}\d X^k(y)\
\mu_{\bar k}^l p_l(z)\ \mu_{\bar j}^r p_r(w)\rangle
\over ({\bar x}-{\bar w})^2({\bar y}-{\bar z})^2}\ \equiv\ F(x,y;z,w)+F(x,y;w,z)
\ee
The straightforward computation of \rf{gmmm} (which again uses only the operator product expansions
of the ``fundamental'' world-sheet fields \rf{px} with their complex conjugated, and integration
by parts over the zero modes,
like in \rf{tarint}) gives rise now to the result
\be
\label{gmmmr}
F(x,y;z,w)
 = {g^{i{\bar j}}\over ({\bar x}-{\bar z})^2({\bar y}-{\bar w})^2}\left(
{\mu_{\bar j}^l\left(\mu^k_{\bar k}\d_k\d_{[l}{\bar\mu}^{\bar k}_{i]}+
\d_i\mu^k_{\bar k}\d_{[l}{\bar\mu}^{\bar k}_{k]}\right)
\over (x-z)(x-w)(y-z)(y-w)}\right. +
\\
+\left. {\d_{[l}{\bar\mu}^{\bar k}_{i]}\mu_{\bar k}^k\d_k\mu^l_{\bar j}
\over (x-y)(x-w)(y-z)(z-w)} +
{\d_{[i}{\bar\mu}^{\bar k}_{l]}\mu_{\bar j}^k\d_k\mu^l_{\bar k}
\over (x-y)(x-z)(y-w)(z-w)} \right. +
\\
+ \left. {\d_i\mu^k_{\bar k}\d_{k}\left({\bar\mu}^{\bar k}_{l}\mu_{\bar j}^l\right)
\over (x-w)^2(y-z)^2}
+ {\d_i\mu^k_{\bar j}\d_{k}\left({\bar\mu}^{\bar k}_{l}\mu_{\bar k}^l\right)
\over (x-z)^2(y-w)^2} -
{{\bar\mu}^{\bar k}_{i}\d_k\mu^l_{\bar j}\d_{l}\mu_{\bar k}^k
\over (x-y)^2(z-w)^2}
\right) =
 \\
= \frac{g^{i{\bar j}}}{({\bar x}-{\bar z})^2({\bar y}-{\bar w})^2}\left(
  \frac{B^{(3)}_{i\bar{j}}}{ (x-y)(x-w)(y-z)(z-w)}
+\frac{ \hat{B}^{(3)}_{i\bar{j}}}{ (x-z)(x-w)(y-z)(y-w)}  \right.
\\
+ \left. {\d_i\mu^k_{\bar k}\d_{k}\left({\bar\mu}^{\bar k}_{l}\mu_{\bar j}^l\right)
\over (x-w)^2(y-z)^2}
+ {\d_i\mu^k_{\bar j}\d_{k}\left({\bar\mu}^{\bar k}_{l}\mu_{\bar k}^l\right)
\over (x-z)^2(y-w)^2} -
{{\bar\mu}^{\bar k}_{i}\d_k\mu^l_{\bar j}\d_{l}\mu_{\bar k}^k
\over (x-y)^2(z-w)^2}
\right)
\ee
where the last equality holds due to an identity
\be
\label{hirgam}
{1\over (x-z)(x-w)(y-z)(y-w)} - {1\over (x-y)(x-w)(y-z)(z-w)} +
\\
+ {1\over (x-y)(x-z)(y-w)(z-w)} = 0
\ee
The structure $g^{i{\bar j}}B^{(3)}_{i\bar{j}} =
g^{i{\bar j}}\d_{[l}{\bar\mu}^{\bar k}_{i]}\mu_{\bar k}^k\d_k\mu^l_{\bar j}$ reminds exactly what should arise at third order in the expansion of the squared Kodaira-Spencer equations
\rf{KoSpe}, while the second term in the r.h.s. of \rf{gmmmr} is given by
\be
\label{1zco}
g^{i{\bar j}}{\hat B}^{(3)}_{i{\bar j}} =
g^{i{\bar j}}\mu_{\bar j}^l\left(\mu^k_{\bar k}\d_k\d_{[l}{\bar\mu}^{\bar k}_{i]}+
\d_i\mu^k_{\bar k}\d_{[l}{\bar\mu}^{\bar k}_{k]}
+\d_l\mu^k_{\bar k}\d_{[k}{\bar\mu}^{\bar k}_{i]}\right) =
\\
= g^{i{\bar j}}\mu^l_{\bar j}\d_l\left(\mu^k_{\bar k}\d_{[k}{\bar\mu}^{\bar k}_{i]}\right)
+ g^{i{\bar j}}\d_i \mu^l_{\bar j}\left(\mu^k_{\bar k}\d_{[k}{\bar\mu}^{\bar k}_{l]}\right)
= -g^{i{\bar j}}\left(\mu^l_{\bar j}\d_l {\sf w}_i +
{\sf w}_l\d_i \mu^l_{\bar j}\right) =
\\
= {\sf w}_i\left(\d_kg^{i{\bar j}}\mu^k_{\bar j}-g^{k{\bar j}}\d_k\mu^i_{\bar j} \right) =
{\sf w}_i{\sf v}^i
\ee
for (cf. with the formulas \rf{bmbm} and \rf{vgm} for OPE's, see Appendix~\ref{ap:ope})
\be
\label{vw}
{\sf v}^i = \mu^k_{\bar k}\d_k g^{i{\bar k}} - \d_k \mu^i_{\bar k}g^{k{\bar k}}
\\
{\sf w}_i = \d_{[i}\bar{\mu}^{\bar{k}}_{k]}\mu^k_{\bar{k}}
\ee
valid modulo transversality constraints \rf{transv} and target-space integration by parts \rf{tarint}.

\section{Divergences in the 4-point function
\label{ss:div4pt}}

Only the coefficients at the terms with $g^{i{\bar j}}B^{(3)}_{i\bar{j}}$ and
$g^{i{\bar j}}\hat{B}^{(3)}_{i\bar{j}}={\sf w}_i{\sf v}^i$
can give rise to the logarithmic divergences, while other are either finite or divergent as
powers. There is a lot of arguments, why the power divergences can be thrown away from many
different angles of view (see e.g. \cite{Pol,Cardy}). We are not going to discuss this now, and
turn instead directly to the logarithmically divergent integrals.

The naive direct calculation shows (see details in Appendix~\ref{ap:int}) that
the logarithmically divergent contribution has the following form
\be
\label{gfi4cft}
C_3 = \frac{1}{3!}\left(\prod_{i=1}^3\int_{\Sigma} {d^2z_i\over\pi}\right)
\langle O_g(x)\Phi(z_1)\Phi(z_2)\Phi(z_3)\rangle =
\\
=\frac{1}{3!}\int_{\Sigma^{\otimes 3}} {d^2y\over\pi}{d^2z\over\pi}{d^2w\over\pi}\left(
\frac{g^{i{\bar j}}B^{(3)}_{i\bar{j}}}
{({\bar x}-{\bar z})^2({\bar y}-{\bar w})^2 (x-y)(x-w)(y-z)(z-w)}+\right.
\\
\left.+\frac{ {\sf w}_i{\sf v}^i}
{({\bar x}-{\bar z})^2({\bar y}-{\bar w})^2 (x-z)(x-w)(y-z)(y-w)}\right)
\\
\approx \frac{2g^{i\bar{j}}B^{(3)}_{i\bar{j}} +
{\sf w}_i{\sf v}^i}{|x|^4}\log \frac{R^2}{\epsilon^2}\int\limits_{|y|<R} \frac{d^2y}{\pi} +c.c
\ee
and comes from the two first terms in the r.h.s. of \rf{gmmmr}. The first part of this
contribution looks exactly as an expected cubic piece
of the squared Kodaira-Spencer equations \rf{KoSpe}. However, the naive computation also shows
that there exists also a logarithmically divergent piece, proportional
to the pairing ${\sf v}^i{\sf w}_i$, with ${\sf v}$ and ${\sf w}$ from \rf{vw}. Let us
now analyze the origin of these two contributions in detail.

Remember first, that the integrals in \rf{gfi4cft} are divergent and should be properly regularized,
  as in sect.~\ref{ss:3pt}, both in the UV and IR regimes. The result for the correlation
  function \rf{gfi4cft} contains {\em all} possible divergences, arising in this way.
  Since the role of the probe operator $O_g(x)$ is
  completely different from the rest when studying the beta-functions, in order to get the latter
  one should pay attention to the structure
  of potential divergences at $|x|\to\infty$.
  It can be easily found, that two logarithmically divergent integrals behave
  quite in a different way.

\begin{figure}[tp]
\epsfysize=4cm
\centerline{\epsfbox{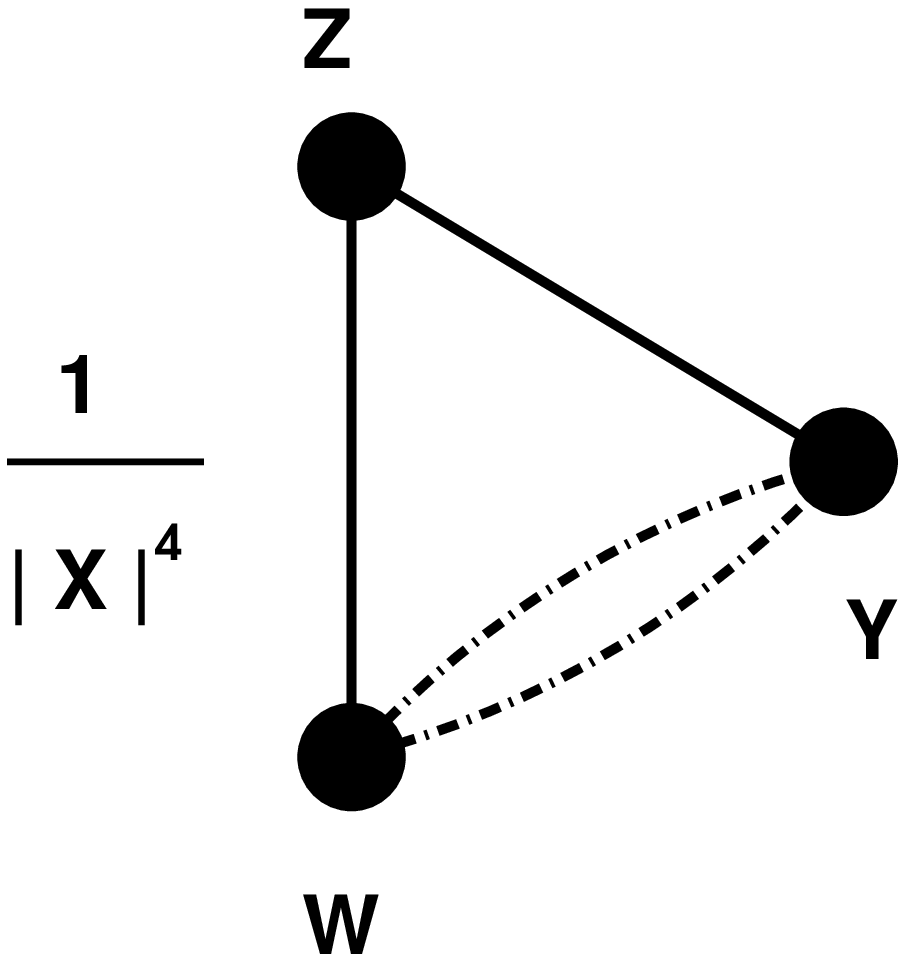}}
\caption{The $|x|\to\infty$ limit ${1\over |x|^4}I_3$ of the first integrand in \rf{gfi4cft}
gives rise to a triangle loop, producing the logarithmic divergency.}
\label{fi:4pt1l}
\end{figure}
The first integrand
\be
\label{triint}
I_3=\int\int{d^2 z\over\pi}{d^2 w\over\pi}{1\over (z-y)(w-z)({\bar w}-{\bar y})^2}
\sim \int{d^2 z\over\pi}{1\over |z-y|^2}
\sim \log{R^2\over\epsilon^2}
\ee
proportional to the desired contribution $g^{i\bar{j}}B^{(3)}_{i\bar{j}}$,
can be described by a triangle loop (in configuration space!)
(see fig.~\ref{fi:4pt1l}), which
being integrated itself diverges logarithmically.
\begin{figure}[tp]
\epsfysize=2cm
\centerline{\epsfbox{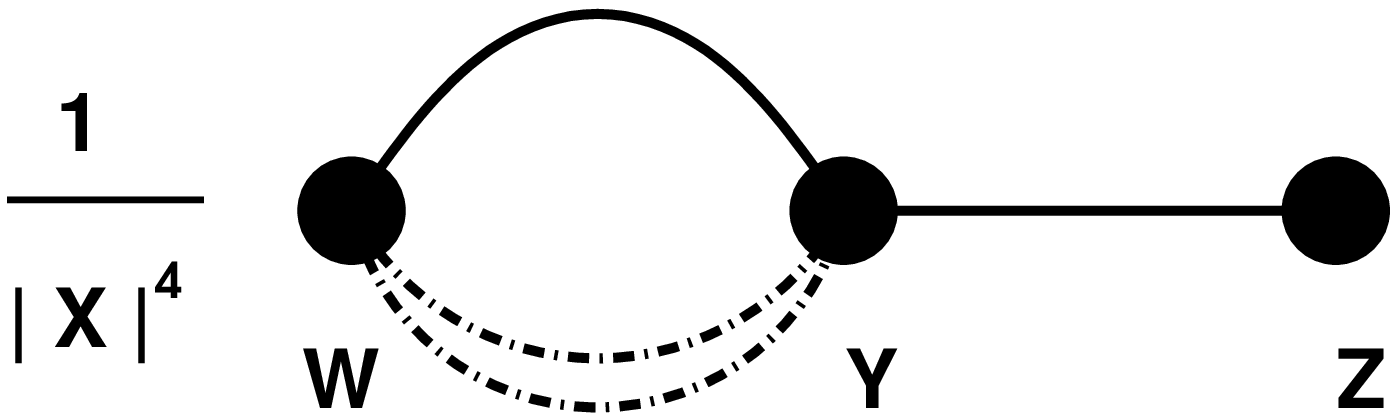}}
\caption{The tadpole ${1\over |x|^4}{\hat I}_3$, arising in the $|x|\to\infty$
limit of the second integrand in \rf{gfi4cft}.}
\label{fi:4pt4l}
\end{figure}
However, this is not the case of the second integrand, whose form in this limit is depicted at
the fig.~\ref{fi:4pt4l}. The divergent tadpole gives rise to the {\em power}
divergency, to be killed after the angle integration as itself, but producing later
the less (logarithmically) divergent contribution
\be
\label{tadint}
{\hat I}_3 = \int{d^2 z\over\pi}{1\over z-y}\int{d^2 w\over\pi}
{1\over (w-y)({\bar w}-{\bar y})^2}
\sim \log{R^2\over\epsilon^2}
\ee
when coupled to the rest of the integrand in \rf{tadint}.

More detailed analysis of the origin of these divergences can be found in
Appendix~\ref{ap:uvir}. The logarithmic divergence of $I_3$ can be separated from
that of ${\hat I}_3$ when carefully analyzing the integration
domains, leading to these divergences. A direct
computation in regularized theory shows, that the desired contribution into $I_3$ is
saturated by UV domain, while the logarithmic divergence of ${\hat I}_3$ really comes from
the IR region. Of course, these words should be pronounced themselves with great care: since
we are dealing with conformal theory the UV and IR domains are related by modular transformations, and therefore when studying the correlation functions
it is not possible to distinct strictly these two
sources of singularity. We have observed already, that these two divergences are in fact mixed
by the identity \rf{hirgam}, which is however violated in the regularized theory.

All these arguments indeed suggest, that the first logarithmic singularity $B^{(3)}_{i{\bar j}}$
should be combined with contribution $B^{(2)}_{i{\bar j}}$ from the 3-point function in order
to obtain the exact form of the squared Kodaira-Spencer equation. This divergency on the
other hand is equivalent to the computation of the beta-function of the operator \rf{vb}, as
already discussed in sect.\ref{ss:3pt}. This beta-function has been computed in \cite{GLM1},
studying the logarithmic divergences in the effective action, (see also various issues of
this procedure e.g. in \cite{FT,GSW,ZJ}).

Decomposing the would-sheet fields into the fast and slow (or quantum and classical) parts
$X \to X_{\rm cl}+ \sqrt{\alpha'}X$ and $p \to p^{\rm cl} + \sqrt{\alpha'}p$, and
expanding the perturbed Lagrangian \rf{act} up to the second order one gets
\be
\label{Lexp}
{\cal L} = {\cal L}_0 + \Phi
= {\cal L}_{\rm cl} + \alpha'\left( p_i {\bar \d}{\tilde X}^i +
p_{i}U^{i}_{\bar{j}}{\tilde X}^{\bar{j}}+
p_{i}W^{i}_{j}{\tilde X}^{j} + c.c\right) + o(\alpha')
\ee
where ${\tilde X}^i = X^i + \mu^i_{\bar k}(X_{\rm cl})X^{\bar k}$, and
\be
\label{Mmatr}
M^i_j = \left(\delta^i_j-(\mu\bar\mu)^i_j\right)^{-1}
\ee
The vertices
\be
\label{vertV}
U^{i}_{\bar{j}} = N_{{\bar k}{\bar l}}^i
{\bar M}^{\bar k}_{\bar j}(X_{\rm cl}){\bar \d}X_{\rm cl}^{\bar l}
\ee
(where the components of the Nijenhuis tensor are defined in \rf{KoSpe})
lead to the only logarithmically divergent contribution in the effective action
\be
\label{divdi}
\Gamma^{\rm div} \sim U^{i}_{\bar{j}}\bar{U}_{i}^{\bar{j}}\int {d^2q\over q{\bar q}}
\ee
The logarithmically divergent integrals in \rf{divdi} lead to renormalization of the operator \rf{vb}
\be
\delta b_{i\bar{j}} \sim \log\epsilon \cdot B_{i\bar{j}}
\ee
where the beta-function of the $b$-field is
\be\label{betab}
B_{i{\bar j}} =
-N^l_{{\bar k}{\bar j}}{\bar N}^{\bar l}_{ki}M^k_l{\bar M}^{\bar k}_{\bar l}
\equiv B^{(2)}_{i{\bar j}} +B^{(3)}_{i{\bar j}}+\bar{B}^{(3)}_{i{\bar j}} + O(\mu^4) =
\\
=-\d_{[k}{\bar\mu}^{\bar k}_{i]}\d_{[\bar{k}} \mu^k_{\bar{j}]}
 + \d_{[k}{\bar\mu}^{\bar k}_{i]}\mu^l_{[\bar{k}}\d_l\mu^i_{\bar{j}]}
+ \d_{[\bar{k}} \mu^k_{\bar{j}]}\bar{\mu}^{\bar{l}}_{[k}\d_{\bar{l}}\bar{\mu}^{\bar{k}}_{i]}+
O(\mu^4)
\ee
and obviously vanishes on solutions to \rf{KoSpe}. Expansion in the equation \rf{betab} contains
explicitly the pieces $B^{(2)}_{i{\bar j}}$ and $B^{(3)}_{i{\bar j}}$, computed above as
target-space coefficients in front of logarithmically divergent parts of 3-point and 4-point correlation functions respectively. The result \rf{betab} is an exact one-loop beta function
\cite{GLM1}, since there are no higher-loop contributions, when computed by the background
field method. It does not mean, that the result should be identically the same in any
other scheme of calculations, and it does not imply certainly, that all divergences of
the correlation function \rf{gfi4cft} are reduced to \rf{betab}.

Hence, let us finally point out, how the ``wrong divergence'' \rf{wrongdiv} can be
possible seen in the effective action. The vertices
\be
\label{vertW}
W^{i}_{j}={\bar \d}X_{\rm cl}^{\bar j}\left(\d_j\mu^i_{\bar j}(X_{\rm cl})
- N_{{\bar k}{\bar j}}^i{\bar\mu}^{\bar k}_s
M^s_j(X_{\rm cl}) \right)
\ee
in \rf{Lexp} could be perhaps neglected,
if the current $j_v$ from \rf{cur} is not anomalous: they can produce only the
linearly divergent
tadpole diagrams, where the divergency is killed by the angle integration.
However, since ${\bar\d}\langle j_v\rangle =
{1\over 2\pi}R^{(2)}\d_iv^i$, the presence of vertices \rf{vertW} in the Lagrangian leads to creation
of the terms
\be
\label{anoct}
W^{i}_{j}\langle p_{i}{\tilde X}^{j}\rangle \sim
{\bar \d}X_{\rm cl}^{\bar j} N_{{\bar k}{\bar j}}^i{\bar\mu}^{\bar k}_s
M^s_j(X_{\rm cl}) = {\bar \d}X_{\rm cl}^{\bar j}
N_{{\bar k}{\bar j}}^i{\bar\mu}^{\bar k}_i(X_{\rm cl}) + O(\mu^4)
\ee
These are the extra terms to be related to the higher
singularities in the operator product expansions and giving rise to extra singularities
in the correlation functions.

It is not amusing also, that the coefficient at \rf{tadint} is
proportional to the higher-order singularity in the operator product expansion \rf{OOPE}
of two $\Phi$ operators (see Appendix \ref{ap:ope} for details). In fact, it is coupled
with a similar higher-order singularity in the OPE of $\Phi$ with
$O_g$ \rf{fiog}, \rf{vgm}. The detailed discussion of the role of such terms goes beyond the scope of
this paper, but we would like to note, that they are directly related with the ``anomalous''
contributions \rf{anoct} to the effective action. Of course, in this order (quadratic in
operators $\Phi$) one gets only the linear part of the full Nijenhuis tensor \rf{KoSpe},
coupled to $dX_{\rm cl}$ in \rf{anoct}.

Indeed, an important fact is that one can rewrite
\be
{\sf w}_i = \d_{[i}\bar{\mu}^{\bar{k}}_{k]}\mu^k_{\bar{k}} = {\bar N}_{ik}^{\bar k}\mu^k_{\bar{k}}
+ O(\mu^3)
\ee
i.e. the extra divergency is still proportional to the Kodaira-Spencer equations \rf{KoSpe} up to the cubic in $\mu$ terms. Therefore, this extra contribution
\be
\label{wrongdiv}
{\sf v}^i{\sf w}_i = {\sf v}^i{\bar N}_{ik}^{\bar k}\mu^k_{\bar{k}} + O(\mu^4)
\ee
still vanishes on the Kodaira-Spencer equations modulo quartic in $\mu$ terms. These quartic
in $\mu$ terms follow only from the 5-point function
$\langle O_g(x)\left(\int_\Sigma d^2z \Phi(z)\right)^4\rangle$. It is easy to see,
that relevant structures indeed appear during its computation, though the exact analysis of the
5-point function (which certainly contains more divergent contributions) goes beyond the scope
of this paper.

\section{BRST approach}

It has been already proposed in \cite{LMZ,GLM1} that the background equations of motion can be
encoded within the BRST approach by a sort of generalized Maurer-Cartan equation
\be
\label{mc}
{\cal Q}_{BRST}(V) \equiv Q_{BRST}V + m_2(V,V) + m_3(V,V,V) + \ldots = 0
\ee
where $m_n(V,\ldots,V)$ stay for certain polyvertex structures, related directly
to the multipoint correlation functions, partially have been computed
above. This constraint can be thought as nonlinear deformation of the initial BRST operator
$Q_{BRST}=Q+{\bar Q}$, defined in terms of the world-sheet conformal theory
\be
Q = Q_{M}+ Q_{gh} = \int dz\left( cT_M + \frac{1}{2} c T_{gh}\right)
\ee
to be considered together with its complex conjugated, where
\be
T = T_M + T_{gh} = {1\over\alpha'}p_{i}\d X^{i} -2\d c b + c \d b
\ee
is the holomorphic stress-tensor of the first order theory \rf{act} coupled to the
anticommuting $[b_n,c_m]_+ = \delta_{n+m,0}$ ghost system
$b(z)=\sum_k b_kz^{-k-2}$ and $c(z)=\sum c_k z^{-k+1}$ of bosonic string.
If operator $\phi$ that does not contain ghosts has conformal dimension $(\Delta,\bar{\Delta})$ then
\be
Q_{BRST}(c\bar{c}\phi)= \bar{\d} \bar{c} c \bar{c}(\bar{\Delta}-1)\phi +
\d c c \bar{c} (\Delta-1)\phi
\ee
and $Q_{BRST}$ just indicates the marginality of the operator $c\bar{c}\phi$.

The physical spectrum of the first-order theory \rf{act} can be identified with the
BRST cohomologies: equation \rf{mc} in the leading order ($Q_{BRST}V=0$), when considering
the vertex operator as perturbation. For example
\be
Q_{BRST}(c\bar{c}g^{i\bar{j}}p_ip_{\bar{j}}) = \frac{\alpha'}{2}\left(\d^2cc\bar{c}\d_ig^{i\bar{j}}p_{\bar{j}}+
\bar{\d}^2\bar{c}c\bar{c}\d_{\bar{j}}g^{i\bar{j}}p_i\right)
\ee
and
\be
Q_{BRST}(c\bar{c}p_i\mu^{i}_{\bar{j}}\bar\d X^{\bar{j}}) =
\frac{\alpha'}{2}\d^2cc\bar{c}\d_i\mu^i_{\bar{j}}\bar{\d} X^{\bar{j}}
\ee
justify that operators \rf{vg} and \rf{vmu} are primary upon the transversality constraints
\rf{transv}, while $Q_{BRST}(c\bar{c}b_{i\bar{j}}\partial X^i\bar{\partial} X^{\bar{j}})=0$
identically. These operators are defined modulo gauge transformations
$V \sim V + Q_{BRST}U$ with parameters
\be
\label{Ua}
U= - {1\over\alpha'}c \alpha^{i}p_i + {1\over\alpha'}c\beta_i\d X^i + c.c.
\ee
giving rise to the pure gauge or BRST-exact terms
\be
\label{QU}
Q_{BRST}U = -\half c\d^2c\d_i\alpha^i + {1\over\alpha'}
c\bar{c}(\d_{\bar{j}}\alpha^{i}p_i - \d_{\bar{j}}\beta_i\d X^i )\bar{\d} X^{\bar{j}} +
c.c.
\ee
i.e. the background fields $\mu$ and $b$ are defined modulo
\be
\label{gauge}
\mu^i_{\bar j} \sim  \mu^i_{\bar j} + \d_{\bar{j}}\alpha^{i}
\\
b_{i\bar{j}} \sim b_{i\bar{j}} - \d_{\bar{j}}\beta_i - \d_i\beta_{\bar{j}}
\ee
and it also follows from \rf{QU}, that the transversality $\d_i\alpha^i=0$
should be assumed in \rf{Ua}, \rf{gauge}.

Beyond the leading order, one should rather study the full equation \rf{mc}, step
by step in perturbation theory. The second term in \rf{mc} in order to ensure the
nilpotency of the deformed BRST operator ${\cal Q}_{BRST}$ should be consistent
with the Leibnitz rule for undeformed $Q_{BRST}$. This does not fix it uniquely,
but one can at least naively try
\be
\label{m2def}
m_2(U,V) = \ha\oint_{|z|=\epsilon} \left(dz U^{(1,0)}(z) + d{\bar z}U^{(0,1)}(z)\right)V^{(0,0)}(0)
\ee
For example, if $U=V$, with the operator
$\Phi dz\wedge d{\bar z}=V^{(1,1)}$ given by \rf{Fimu}, or
\be
\label{VFi}
V^{(0,0)} =c\bar{c}(p_{i}\mu^{i}_{\bar{j}}\bar{\d} \bar{X}^{\bar{j}}+\bar{p}_{\bar{i}}
\bar{\mu}^{\bar{i}}_{j}\d X^j)
\\
V^{(1,0)} = b_{-1}V^{(0,0)} =\bar{c}(p_{i}\mu^{i}_{\bar{j}}\bar{\d} \bar{X}^{\bar{j}}
+\bar{p}_{\bar{i}}\bar{\mu}^{\bar{i}}_{j}\d X^j)
\ee
formula \rf{m2def} gives rise to
\be
\label{m2calc}
m_2(V,V) = 2(\bar{\d} \bar{c} c \bar{c} +
\d c c \bar{c})\left(\frac{f}{\epsilon^2} + {\tilde B}^{(2)}_{i\bar{j}}\d
X^i\bar{\d} \bar{X}^{\bar{j}} \right)+\frac{1}{2}\d^2 c c\bar{c}
\d_{\bar{j}}f \bar{\d} \bar{X}^{\bar{j}}+\frac{1}{2}\bar{\d}^2
\bar{c} c\bar{c} \d_if \d X^i
\ee
where $f = \bar{\mu}^{\bar{i}}_{j}\mu^{j}_{\bar{i}} = \Tr\ \bar{\mu}\mu$, and
\be
\label{tilB}
{\tilde B}^{(2)}_{i\bar{j}} = -\d_{[k}\bar{\mu}^{\bar{k}}_{i]}\d_{[\bar{k}}
\mu^k_{\bar{j}]}
+ \ha\d_i\left(\bar{\mu}^{\bar{k}}_{k}\d_{[\bar{j}} \mu^{k}_{\bar{k}]}\right)
+ \ha\d_{\bar{j}} \left( \mu^{k}_{\bar{k}}\d_{[i}  \bar{\mu}^{\bar{k}}_{k]}\right)\equiv
\\
\equiv B^{(2)}_{i\bar{j}} + \half\left(\d_i{\sf w}_{\bar j}
+ \d_{\bar j}{\sf w}_i\right)
\ee
Equation \rf{mc} in the next order (where $V^{(2)}$ is deformation of the original
$V=V^{(1)}$)
\be
\label{mc2}
Q_{BRST}V^{(2)} + m_2(V,V) = 0
\ee
shows that the expression \rf{m2calc} can be considered modulo the BRST-exact terms.
Choosing $V^{(2)}$ in the form
\be
\label{V2}
V^{(2)} = c\bar{c}\frac{f}{\e^2} +
 \left(\frac{c\bar{c}}{\alpha'}p_{i}\mu^{(2)i}_{\bar{j}}\bar{\d} X^{\bar{j}}
 +\d^2cc \frac{f}{4}
+ \half c(\d c +\bar{\d} \bar{c}){\sf w}_i \d X^{i} +c.c.\right)
\ee
and substituting it together with \rf{m2calc} into \rf{mc2}, one gets
\be\label{mc2calc1}
Q_{BRST}V^{(2)} + m_2(V,V) =
\bar{\d} \bar{c} c \bar{c}B^{(2)}_{i\bar{j}} \d X^{i}
\bar{\d} X^{\bar{j}} + \frac{1}{2}\d^2 cc \bar{c}
( \d_{i}\mu^{(2)i}_{\bar{j}}-{\sf w}_{\bar{j}})\bar{\d} X^{\bar{j}} + c.c. = 0
\ee
so that it requires the vanishing of the beta-function \rf{betab} at given order
$B^{(2)}_{i\bar{j}} =0 $ together with
${\sf w}_{\bar{j}} = \d_{[{\bar j}}{\mu}_{\bar{k}]}^{k}{\bar \mu}_k^{\bar{k}} =
\d_{i}\mu^{(2)i}_{\bar{j}}$. Due to the first condition
$\d_{[\bar{j}}\mu^i_{\bar{k}]} = 0$ (together with its complex conjugated) and therefore
$\mu^{(2)}$ in \rf{V2} can be chosen trivial.

Equation \rf{mc2} itself is defined modulo the symmetry transformation
\be
\label{sym2}
V^{(2)} \rightarrow V^{(2)}  + Q_{BRST}U^{(2)} - m_{2}(V,U) - m_{2}(U,V)
\ee
where it is convenient to choose $U^{(2)}$ in the following form
\be
U^{(2)} =  -\frac{c}{\alpha'} \left(\alpha^{(2)i}-\half\alpha^{\bar{k}}\mu^{i}_{\bar{k}}\right)p_i
  + \frac{\bar{c}}{\alpha'} \left(\alpha^{(2)\bar{j}}-\half\alpha^{k}\bar{\mu}^{\bar{j}}_{k}
  \right)\bar{p}_{\bar{j}}
\ee
Then the transformation \rf{sym2} leads in particular to
\be\label{sym2calc}
\mu^{(2)i}_{\bar{j}} \sim \mu^{(2)i}_{\bar{j}}  +\d_{\bar{j}}(\alpha^{(2)i}-\alpha^{\bar{k}}
\mu^{i}_{\bar{k}})+\alpha^{k}\d_{k}\mu^{i}_{\bar{j}} + \alpha^{\bar{k}}\d_{\bar{k}}\mu^{i}_{\bar{j}}
- \mu^{k}_{\bar{j}}\d_k\alpha^{i} + \mu^{i}_{\bar{k}}\d_{\bar{j}}\alpha^{\bar{k}}
\ee
which completes (in a given order of perturbation expansion)
the gauge transformation of the Beltrami fields from \rf{gauge}
to the full nonlinear form (see \cite{GLM1})
\be\label{mutr}
\mu^{i}_{\bar{j}} \rightarrow \mu^{i}_{\bar{j}} +
\d_{\bar{j}}v^i + v^{k}\d_k\mu^{i}_{\bar{j}} +
v^{\bar{k}}\d_{\bar{k}}\mu^{i}_{\bar{j}}
+\mu^{i}_{\bar{k}}\d_{\bar{j}}v^{\bar{k}} -
\mu^k_{\bar{j}}\d_kv^i
 - \mu^i_{\bar{k}}\mu^k_{\bar{j}}\d_k v^{\bar{k}} - g^{i{\bar k}}b_{l{\bar j}}\d_{\bar k}v^l=
\\
=\mu^{i}_{\bar{j}} + \d_{\bar{j}}v^i + \{ v, \mu \}^i_{\bar j} +
v^{\bar{k}}\d_{\bar{k}}\mu^{i}_{\bar{j}} +\mu^{i}_{\bar{k}}\left(\d_{\bar{j}}
- \mu^k_{\bar{j}}\d_k\right) v^{\bar{k}} - g^{i{\bar k}}b_{l{\bar j}}\d_{\bar k}v^l
\ee
upon the identification
\be
\label{valfa}
v^{(2)i} = \alpha^{(2)i}-\alpha^{\bar{k}}\mu^i_{\bar{k}}
\ee
This formula with already mentioned transversality condition $\d_{i}\alpha^{(2)i} =0$ for the $U$-fields
is also consistent with the transversality constraint for parameter $v$ of the full gauge
transformation.
The exact form of the symmetry transformation \rf{mutr} can be restored studying all
higher operations from the equation \rf{mc}.

\section{Discussion}

In this paper we have computed the correlation functions in
the first-order conformal field theory, which hopefully has a sense of expansion
of bosonic string theory in singular backgrounds.
This theory depends only on the (local) choice of target-space complex structure and we have
studied this dependence by perturbing the bare action with the Beltrami vertex operator
\rf{Fimu}, and computing its correlations with the ``probe'' operator \rf{vg}.

We have derived the Kodaira-Spencer equations \rf{KoSpe} for the target-space
Beltrami differentials from computation of the correlation functions on punctured spheres.
The ``co-ordinate'' beta-functions arise in such approach as certain integrals
over the moduli spaces of punctured world sheets of the first-order string theory. We have
demonstrated how the contribution to the squared Kodaira-Spencer equations comes from
the 3-point and 4-point correlators for two or three operators \rf{Fimu} correspondingly and
a single spectator ``operator'' \rf{vg}.

We have studied the logarithmic divergences, arising in these correlation functions.
For the 3-point function this divergency immediately reproduces the quadratic contribution
to the beta-function of the dual to probe operator \rf{vb}, being essentially equivalent
to the computation of this contribution from OPE's of the perturbing fields \cite{Pol,Zam}.
This beta-function has the form of the {\em squared} Kodaira-Spencer equations, and the
computation of the 3-point function reproduced it in the first nonvanishing order. In order
to complete this derivation, and to get full Kodaira-Spencer equation, as required for example
by target-space symmetries of the theory, we have computed the 4-point function, again with
one operator being the probe operator \rf{vg}. The divergency of the 4-point function gives rise
to a necessary contribution to the squared Kodaira-Spencer equations \rf{betab}, which
is cubic in Beltrami fields and completes their derivation in given order. We have also analyzed
an extra divergency in the 4-point function, and demonstrated that it origin comes rather
from IR, than from UV regime, which however could be hardly distinguished in conformal
field theory. Nevertheless, in given order this extra divergency also vanishes on Kodaira-Spencer equations \rf{KoSpe}.

Since the nonlinear equations for the
background fields arise in this approach from the correlation functions with different
number of operators (the so called polyvertex structures, already discussed in this context
in \cite{LMZ,Zeitl}),
and contain integrals over the moduli spaces of punctured world sheets with different numbers
of punctures, it is natural to understand them also in the context of BRST approach.
This is close to already discussed in similar context structures of the
string field theory (see e.g. \cite{Zwi,Son}). We have proposed how the first nontrivial
polyvertex structure can look like, which turns the condition of BRST-closeness into nonlinear
generalized Maurer-Cartan equation \rf{mc}. We have analyzed this equation in the first
nontrivial order and demonstrated that its analysis is consistent with other approaches.
Generally one may even expect some solvable nonlinear equations for
the ``co-ordinate'' beta-functions, arising when studying the Maurer-Cartan equation, and we are planning to return to these problems elsewhere.

\bigskip\noindent
{\bf Acknowledgements}

\noindent
We are grateful to G.~Arutyunov, M.~Kontsevich, A.~S.~Losev, N.~Nekrasov, A.~Rosly, A.~Smilga
and A.~Tseytlin for important discussions. OG also thanks V.~Shadura and N.~Iorgov
for permanent encouragement and support. AM thanks the IHES in Bures-sur-Yvette,
the MPIM in Bonn and the Galileo Galilei Institute in Florence, where essential
parts of this work have been done.

The work of OG was supported by the grant for support of Scientific Schools
LSS-3036.2008.2, by French-Ukrainian program ``Dnipro'', the
project  M17-2009, the joint project PICS of CNRS and Nat. Acad. Sci. of Ukraine and the project
F28.2/083 of FRSF of Ukraine.
The work of AM was supported by the Russian Federal Nuclear Energy Agency,
the RFBR grant 08-01-00667,
the grant for support of Scientific Schools LSS-1615.2008.2, the
INTAS grant 05-1000008-7865, the project ANR-05-BLAN-0029-01, the
NWO-RFBR program 047.017.2004.015, the Russian-Italian RFBR program 09-01-92440-CE,
the Russian-Ukrainian RFBR program 09-02-90493, the Russian-French RFBR-CNRS program
09-02-93105 and by the Dynasty foundation.

\def\theequation{\thesection.\arabic{equation}}
\section*{Appendix}
\appendix
\def\theequation{\thesection.\arabic{equation}}
\setcounter{equation}0

\section{Operator product expansions
\label{ap:ope}}

In the free theory with the first-order action \rf{act} all computations of the correlation
functions can be performed using the operator product expansions \rf{px} of the basic
fields. From the point of view of renormalization theory \cite{Pol,Zam} the computation
of the operator product expansions of the primary operators corresponds exactly to
the quadratic in coupling terms in the beta functions, expressed through the structure
constants of the operator algebra. As was stated in \cite{LMZ}, the computation of the
operator product expansion of two operators \rf{vg} leads to the beta function of the
operator \rf{vg} itself and requiring this beta-function to vanish, or equivalently
the operator \rf{vg} to be exactly marginal, one gets the nonlinear
equation \rf{geq} for the ``metric'' components $g^{i{\bar j}}$.

For the purposes of this paper let us consider the operator product expansions of
$O_\mu = p_i\mu^i_{\bar i}{\bar\d}X^{\bar i}$ and $O_{\bar\mu} = p_{\bar j}{\bar\mu}^{\bar j}_j\d X^j$.
We shall always keep only the terms with maximally two target-space derivative, since the other
ones will be suppressed in $\alpha'$.
Expanding at $z \to 0$ one gets
\be
\label{opembm}
O_\mu(z)O_{\bar\mu}(0) \sim {{\bar\mu}_i^{\bar j}\mu^i_{\bar j}\over |z|^4} +
{\d X^k\over z{\bar z}^2}\left(\d_k({\bar\mu}_i^{\bar j}\mu^i_{\bar j}) -
{\sf w}_k\right)+
\\
+ {{\bar\d} X^{\bar k}\over z^2{\bar z}}{\sf w}_{\bar{k}} +
{\d X^k{\bar\d} X^{\bar k}\over |z|^2} \left(\d_k({\bar\mu}^{\bar j}_i
\d_{[{\bar k}}\mu^i_{{\bar j}]})+B^{(2)}_{k{\bar k}}\right)+\ldots
\ee
with
\be
\label{bmbm}
B^{(2)}_{k{\bar k}} = \d_{[i}{\bar\mu}^{\bar j}_{k]}\d_{[{\bar k}}\mu^i_{{\bar j}]}
\\
{\sf w}_{\bar{j}}= \d_{[\bar{j}}\mu^k_{\bar{k}]}\bar{\mu}^{\bar{k}}_{k},\,\,\,\,
    {\sf w}_i = \d_{[i}\bar{\mu}^{\bar{k}}_{k]}\mu^k_{\bar{k}}
\ee
while
\be
\label{opemm}
O_\mu(z)O_\mu(0) \sim
- {{\bar\d} X^{\bar i}{\bar\d} X^{\bar j}\over z^2}\ \d_i\mu^j_{\bar j}\d_j\mu^i_{\bar i}\
+\ {p_i{\bar\d} X^{\bar i}{\bar\d} X^{\bar j}\over z}\ \mu^j_{[{\bar i}}\d_j\mu^i_{{\bar j}]} +
+\ldots
\ee
We see that the last formula does not contain at all the ``potentially logarithmic'' non holomorphic terms
$1/|z|^2$, so the properties of the product of two operators \rf{Fimu} will be totally
determined by \rf{bmbm}. The most singular contribution in \rf{bmbm} defines the trace of
Zamolodchikov metric $f=\Tr\ {\bar\mu}\mu = {\bar\mu}_i^{\bar j}\mu^i_{\bar j}$ (which in this
or that way drops out from the interesting formulas), while the term at $1/|z|^2$ (up
to the total derivative!) gives rise to the $O(|\mu|^2)$ contribution to the beta function
of the $b$-field or \rf{vb} operator.

It follows from \rf{opembm} and \rf{opemm} that the
OPE of the operator \rf{Fimu} with itself acquires the following form
  \be
  \label{OOPE}
  \Phi(z)\Phi(0) = {2{\bar\mu}_i^{\bar j}\mu^i_{\bar j}\over |z|^4} +
  {\d_k\left({\bar\mu}_i^{\bar j}\mu^i_{\bar j}\right)\d X^k\over z{\bar z}^2} +
  {\d_{\bar k}\left({\bar\mu}_i^{\bar j}\mu^i_{\bar j}\right){\bar\d} X^{\bar k}\over z^2{\bar z}}
  + \frac{2 {\tilde B}^{(2)}_{i\bar{j}}\d X^i\bar{\d} X^{\bar{j}}}{|z|^2} + \ldots
\\
{\tilde B}^{(2)}_{i\bar{j}} = B^{(2)}_{i\bar{j}} +
\half\left(\d_i{\sf w}_{\bar{j}} + \d_{\bar{j}}{\sf w}_i\right)
   \ee
while the OPE of $\Phi$ with $O_{g}$ \rf{vg} is
\be
\label{fiog}
\Phi(z) O_{g}(0) = {p_i{\sf v}^i\over z{\bar z}^2} + {p_{\bar i}{\sf v}^{\bar i}\over z^2{\bar z}} +
\frac{1}{|z|^2} \left( p_{i} f^{i}_{\bar{d}}\bar{\d} \bar{X}^{\bar{d}}  +
\bar{p}_{\bar{i}} \bar{f}^{\bar{i}}_{d}\d X^{d}
\right) + \ldots
\ee
where
\be
\label{vgm}
{\sf v}^i = \mu^k_{\bar k}\d_k g^{i{\bar k}} - \d_k \mu^i_{\bar k}g^{k{\bar k}}
\\
f^{i}_{\bar{j}} = \d_{[\bar{j}}\mu^k_{\bar{l}]}\d_k g^{i\bar{l}} -
g^{k\bar{l}}\d_k\d_{[\bar{j}}\mu^i_{\bar{l}]}
\ee
is contribution, together with its complex conjugated, to the renormalization of the operators
\rf{vmu}, while
renormalization of the operator \rf{vb} is suppressed in $\alpha'$. Finally, the OPE
\be
\label{gb}
O_g(z)O_b(0) = {g^{i{\bar j}}b_{i{\bar j}}\over |z|^4} +
{g^{i{\bar j}}\d_{\bar j}b_{i{\bar l}}\bar\d X^{\bar l}\over z^2{\bar z}} +
{g^{i{\bar j}}\d_ib_{k{\bar j}}\d X^k\over z{\bar z}^2} +
\\
+ {1\over |z|^2}\left(\d_k\d_{\bar l}g^{i{\bar j}}b_{i{\bar j}} +
g^{i{\bar j}}\d_i\d_{\bar j}b_{k{\bar l}} +
\d_k g^{i{\bar j}}\d_{\bar j}b_{i{\bar l}} +
\d_{\bar l}g^{i{\bar j}}\d_i b_{k{\bar j}}\right)\d X^k\bar\d X^{\bar l} + \ldots
\ee
which we have not been interested in at vanishing bare $b_{i{\bar j}}=0$.

\setcounter{equation}0
\section{Calculation of the integrals
\label{ap:int}}

Let us start with the integral, arising during the computation
of both 3-point and 4-point functions \rf{gfifi} and \rf{gfi4cft}
\be
\label{iint}
I(x-y) = \int_{D_\epsilon(x,y)} {d^2z\over \pi}{1\over |z-x|^2|z-y|^2} =
{i\over 2\pi}{1\over |x-y|^2}
\int_{D_\epsilon(x,y)} d\log{z-x\over z-y}\wedge
d\log{{\bar z}-{\bar x}\over {\bar z}-{\bar y}}
\ee
By Stokes theorem it reduces to the boundary integral
\be
I(x-y) = {i\over 2\pi}{1\over |x-y|^2}
\oint_{\d D_\epsilon(x,y)} \log{{\bar z}-{\bar x}\over {\bar z}-{\bar y}}\ d\log{z-x\over z-y}
\ee
where the contour $\d D_\epsilon(x,y)$, surrounding the cut between the points
$x$ and $y$ is passed
counter-clockwise. The integral therefore reduces to
\be
\oint_{C_-+C_x+C_++C_y} \log{{\bar z}-{\bar x}\over {\bar z}-{\bar y}}\ d\log{z-x\over z-y}=
\\
= 2\pi i\int_{y+\epsilon}^{x-\epsilon} d\log{z-x\over z-y} +
\int_{-\pi}^{\pi}\log{\epsilon e^{-i\phi}\over {\bar x}-{\bar y}}\ id\phi -
\int_{0}^{2\pi}\log{{\bar y}-{\bar x}\over \epsilon e^{-i\phi}}\ id\phi =
\\
= 2\pi i\left(2\log {\epsilon\over y-x}+2\log{\epsilon\over {\bar y}-{\bar x}}\right)
= - 2\pi i\cdot 2\log{|x-y|^2\over\epsilon^2}
\ee
so that one gets for \rf{iint}
\be
\label{itwo}
I(x-y) = {2\over |x-y|^2}\log{|x-y|^2\over\epsilon^2},\ \ \ \ |x-y| > \epsilon
\ee
Instead of \rf{iint} one can consider its $\Gamma$-regularization
\be
\label{iintab}
I(\alpha,\beta) = \int {d^2z\over \pi}{1\over |z-x|^{2\beta}|z-y|^{2\alpha}} =
{1\over |x-y|^{2(\alpha+\beta-1)}}\int {d^2\xi\over \pi}{1\over |\xi|^{2\alpha}|\xi-1|^{2\beta}}
\\
\half<\alpha,\beta<1
\ee
The last integral (the Shapiro-Virasoro amplitude) is taken as Gaussian after an obvious substitution
\be
{1\over|\zeta|^{2\gamma}} =
{1\over\Gamma(\gamma)}\int_0^\infty {dt\over t}t^\gamma e^{-t|\zeta|^2}
\ee
and results in
\be
\label{iiabr}
I(\alpha,\beta) =
{1\over |x-y|^{2(\alpha+\beta-1)}}{\Gamma(1-\alpha)\over\Gamma(\alpha)}
{\Gamma(1-\beta)\over\Gamma(\beta)}{\Gamma(\alpha+\beta-1)\over\Gamma(2-\alpha-\beta)}
\\
\half<\alpha,\beta<1
\ee
In the limit $\alpha,\beta\to 1$ one can take
\be
\label{iid}
I(1-\epsilon,1-\delta) =
{1\over |x-y|^{2(1-\epsilon-\delta)}}{\Gamma(\epsilon)\over\Gamma(1-\epsilon)}
{\Gamma(\delta)\over\Gamma(1-\delta)}
{\Gamma(1-\epsilon-\delta)\over\Gamma(\epsilon+\delta)}\ =
\\
\stackreb{\epsilon,\delta\to 0}{=}\ \left({1\over\epsilon}+{1\over\delta}\right){1\over |x-y|^2}
+ \left(2+{\epsilon\over\delta}+{\delta\over\epsilon}\right){1\over |x-y|^2}\log|x-y|^2
+ O(\epsilon,\delta)
\ee
and the result \rf{itwo} is reproduced only upon $\epsilon^2+\delta^2=0$.

Let us also calculate
\be
J(x-y) = \frac{1}{\bar{x}-\bar{y}}\int_{D_\epsilon(x,y)} {d^2z\over \pi}{1\over |z-x|^2(z-y)}\
\stackreb{z=y+(x-y)\left(\xi+\half\right)}{=}
\\
= \frac{1}{|x-y|^2}\int_{D_{\epsilon/|x-y|}\left(\half,-\half\right)}
\frac{d^2\xi}{\pi} \frac{\xi+\half}{\left|\xi-\half\right|^2\left|\xi+\half\right|^2}
\ee
Due to the symmetry of the integration domain, one immediately gets
that this integral is exactly half of the original integral \rf{iint}
\be\label{iint22}
J(x-y) = \ha\frac{1}{|x-y|^2}\int_{D_{\epsilon/|x-y|}\left(\half,-\half\right)}
\frac{d^2\xi}{\pi} \frac{1}{\left|\xi-\half\right|^2\left|\xi+\half\right|^2} =
\frac{1}{|x-y|^2}\log{|x-y|^2\over\epsilon^2}
\ee

\setcounter{equation}0
\section{UV and IR contributions to the logarithmic divergence of the
4-point function
\label{ap:uvir}}

In order to study the divergences carefully, it is useful to rewrite the regularized
integral \rf{triint} (as well as \rf{tadint}) in the form
\be
\label{cc3}
I_3 = \int\limits_{|z_1|<R,|z_2|<R,|z_3|<R}
\prod\limits_{i=1}^3 \frac{d^2z_i}{\pi} \theta(|z_{12}|-\e)\theta(|z_{13}|-\e)
\theta(|z_{23}|-\e)F(z_1,z_2,z_3) + c.c
\ee
and extract divergent pieces by acting on the \rf{cc3}
by the $ -\e \frac{d}{d\e}$
\be
\label{difI}
 -\e \frac{d}{d\e} I_3 = - \left.\e\sum_{i<j} \frac{dI_3}{d\e}\right|_{z_{ij}}
\ee
and leaving only the constant terms as $\e\to0$, where, say
\be
-\left.\e\frac{dI_3}{d\e}\right|_{z_{12}} =
\e\int\limits_{|z_1|<R,|z_2|<R,|z_3|<R}
\prod\limits_{i=1}^3 \frac{d^2z_i}{\pi}\delta(|z_{12}|-\e)\theta(|z_{13}|-\e)\theta(|z_{23}|-\e)
F(z_1,z_2,z_3)
\ee
Introducing new variables
\be
z_3= z,\,\,\,\, z_3 -z_2 =u,\,\,\,  z_3 -z_1 =v
\ee
one obtains
\be
-\e\frac{dC_3}{d\e}\Big|_{z_{12}} =
\e\int\limits_{\Sigma^{*}}  \frac{d^2z}{\pi}\frac{d^2v}{\pi}\frac{d^2u}{\pi}\delta(|u-v|-\e)
F(z-u,z-v,z)
\ee
where
\be
\Sigma^{*}:  \,\,\,\,\, |z-v|<R,\,\,\,|z-u|<R,\,\,\,|z|<R,\,\,\,|u|>\e,\,\,\,|v|>\e.
\ee
and, therefore
\be
-\e\frac{dI_3}{d\e} = \e\int\limits_{\Sigma^{*}}  \frac{d^2z}{\pi}\frac{d^2v}{\pi}
\frac{d^2u}{\pi}\delta(|u-v|-\e)
\left[F(-u,-v,0) +F(-u,0,-v) + F(0,-v,-u)\right]
\ee
for their sum  \rf{difI}.
Since the integrand $F(z_1,z_2,z_3)$ in \rf{triint} (as well as
${\hat F}(z_1,z_2,z_3)$ in \rf{tadint}) is translational invariant $F(z-u,z-v,z)= F(-u,-v,0)$, and
there is also rotational invariance:
\be
F(e^{i\phi}z_1,e^{i\phi}z_2,e^{i\phi}z_3)=F(z_1,z_2,z_3)
\ee
one of the integrations is easily performed. Indeed, from the delta-function constraint one
concludes that
\be
v=u- \e e^{i\phi}
\ee
but using rotational invariance we can put
\be
v=u - \e
\ee
while the angle integral is just $2\pi$. Hence,
\be
-\e\frac{dI_3}{d\e} = 2\e^2\int\limits_{\Sigma_{\e,R}}  \frac{d^2z}{\pi}\frac{d^2u}{\pi}
\left[F(-u,-u+\e,0) +F(-u,0,-u+\e) + F(0,-u+\e,-u)\right]
\ee
where $\Sigma_{\e,R}$ is the domain (after additional change of the variable $z\to u-z$) defined by
the following inequalities
\be
\Sigma_{\e,R}:\,\,\,\, |z-\e|<R,\,\,\,|z|<R,\,\,\,\, |z-u|<R,\,\,\,\, |u|>\e,\,\,\, |u-\e|>\e
\ee
For the integral \rf{triint}
\be
F(z_1,z_2,z_3) = \frac{1}{\overline{z}_{13}^2}\frac{1}{z_{12}}\frac{1}{z_{23}}
\ee
therefore
\begin{figure}[tp]
\epsfysize=8cm
\centerline{\epsfbox{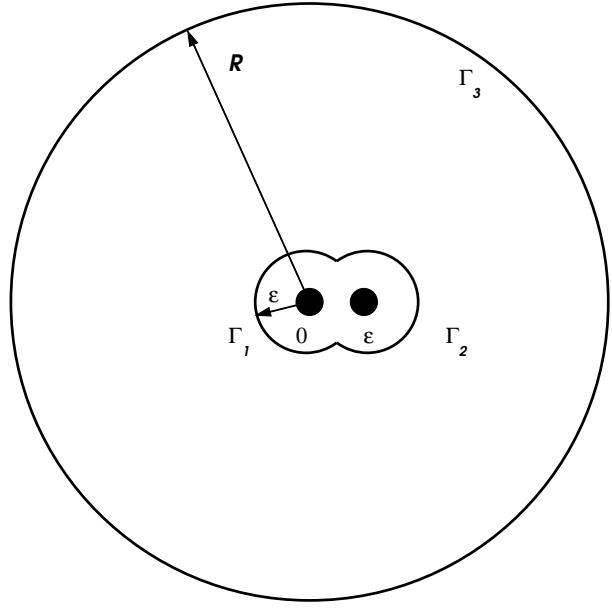}}
\caption{The set of integration contours: $\Gamma_1$ and $\Gamma_2$ for the UV contribution, and
$\Gamma_3$ for the contribution in IR regime.}
\label{fi:cont}
\end{figure}
\be
\e\frac{dI_3}{d\e} = - 2\int\limits_{\Sigma_{\e,R}}
\frac{d^2z}{\pi}\frac{d^2u}{\pi}\left[\frac{2\e}{\overline{u}^2(u-\e)}-\frac{1}{u(u-\e)}
\right] =
\\
= 2\int\limits_{|z|<R,\,|z-\e|<R}  \frac{d^2z}{\pi}
\oint\limits_{\Gamma_{1}\cup\Gamma_{2}\cup \Gamma_{3}}\frac{du}{2\pi i}
\left[\frac{2\e}{\overline{u}(u-\e)}+\frac{\overline{u}}{u(u-\e)}
\right]
\ee
where contours
\be\label{Contour}
\Gamma_1:\,\,\,\,\,\, u= \e e^{i\phi},\, \phi \in \left[\frac{\pi}{3},\frac{5\pi}{3}\right]\,\,\,
\Gamma_2:\,\,\,\,\,\, u= \e+\e e^{i\phi},\, \phi \in \left[\frac{4\pi}{3},\frac{2\pi}{3}\right]\,\,\,\\
\Gamma_3:\,\,\,\,\,\, u= z - R e^{i\phi},\, \phi \in \left[0,2\pi\right]\,\,\,
\ee
are depicted at fig.~\ref{fi:cont}.

An important thing is, that for triangular loop and the integral \rf{triint}, the
integration over $\Gamma_{3}$ or the contribution from the IR domain vanishes
at $\e\to 0$. It means, that the result for the integral \rf{triint} is purely from UV,
and corresponds therefore to the beta-function of the operator \rf{vb}. To compute it, one
first notices, that $\Gamma_2$ is union of two intervals
\be
\Gamma_2 = \left[\frac{4\pi}{3},2\pi\right] \cup \left[0,\frac{2\pi}{3}\right]
\ee
and after the changes of the angle variables $\phi \to \phi +\pi$ on the first interval and
$\phi \to \phi -\pi$ on the second, the integral over $\Gamma_2$ transforms into the integral
over $\Gamma_1$, giving rise to the final result
\be\label{df}
\e\frac{dI_3}{d\e} = 2\int\limits_{|z|<R,\,|z-\e|<R}
 \frac{d^2z}{\pi}
 \int\limits_{\frac{\pi}{3}}^{\frac{5\pi}{3}}\frac{d\phi}{2\pi }
 \frac{2e^{2i\phi}+2e^{i\phi}+2e^{-i\phi}-1}{e^{i\phi}-1}  =
2\int\limits_{|z|<R}  \frac{d^2z}{\pi} + o(\e)
\ee
since
\be
\label{ai}
\int\limits_{\frac{\pi}{3}}^{\frac{5\pi}{3}}\frac{d\phi}{2\pi }
 \frac{2e^{2i\phi}+2e^{i\phi}+2e^{-i\phi}-1}{e^{i\phi}-1}  =
 \left.\left[{2\over i\pi}\cos\phi+
 {5\over 4\pi i}\log 2(\cos\phi-1)+{3\over 4\pi}\phi\right]\right|_{\frac{\pi}{3}}^{\frac{5\pi}{3}}
= 1
\ee
The situation is completely different for the integral \rf{tadint}. In this case
\be
{\hat F}(z_1,z_2,z_3)  =\frac{1}{\overline{z}^2_{13}}\frac{1}{z_{12}z_{13}}
\ee
and therefore
\be
\e\frac{d{\hat I}_3}{d\e} = - 2\int\limits_{\Sigma_{\e,R}}
\frac{d^2z}{\pi}\frac{d^2u}{\pi}\left[\frac{\e}{\overline{u}^2(u-\e)}+\frac{1}{\e u}
\right]=
\\
= 2\int\limits_{|z|<R,\,|z-\e|<R}
\frac{d^2z}{\pi}\oint\limits_{\Gamma_{1}\cup\Gamma_{2}\cup
\Gamma_{3}}\frac{du}{2\pi i}\left[\frac{\e}{\overline{u}(u-\e)}-\frac{\overline{u}}{\e u}
\right]
\ee
Now performing the same trick and taking into account that last term can contribute when we integrate
over the contour $\Gamma_3$ we obtain:
\be
\e\frac{d{\hat I}_3}{d\e} = 2\int\limits_{|z|<R,\,|z-\e|<R}
\frac{d^2z}{\pi} \left(\int\limits_{\frac{\pi}{3}}^{\frac{5\pi}{3}}\frac{d\phi}{2\pi }
\frac{e^{2i\phi}+e^{-i\phi}}{e^{i\phi}-1} -\oint_{\Gamma_3}\frac{du}{2\pi i}
\frac{\overline{u}}{\e u}
\right)
\ee
The first integral in the r.h.s.
\be
\label{angint}
\int\limits_{\frac{\pi}{3}}^{\frac{5\pi}{3}}\frac{d\phi}{2\pi }
\frac{e^{2i\phi}+e^{-i\phi}}{e^{i\phi}-1} =
\left.{1\over i\pi}\left[\cos\phi+\log 2(\cos\phi-1)\right]\right|_{\frac{\pi}{3}}^{\frac{5\pi}{3}}
= 0
\ee
vanishes, while the second -
totally saturated by the IR $\Gamma_3$ contribution - gives rise to the expected result
\be
\e\frac{d{\hat I}_3}{d\e}
= -2\int_{|z|<R,\,|z-\e|<R} \frac{d^2z}{\pi} \frac{\overline{z}}{\e} =
\int\limits_{|z|<R}  \frac{d^2z}{\pi}+ o(\e)
\ee
or the expected half of the result (\ref{df}), though we have seen that the nature of the
divergence in ${\hat I}_3$ is totally different from that of $I_3$.


\end{document}

\end{document}